# Weighted Hypernetworks

Jin-Li Guo[*]　Xin-Yun Zhu

*Business School, University of Shanghai for Science and Technology,*

*Shanghai, 200093, China*

**Abstract:** Complex network theory has been used to study complex systems. However, many real-life systems involve multiple kinds of objects . They can't be described by simple graphs. In order to provide complete information of these systems, we extend the concept of evolving models of complex networks to hypernetworks. In this work, we firstly propose a non-uniform hypernetwork model with attractiveness, and obtain the stationary average hyperdegree distribution of the non-uniform hypernetwork. Furthermore, we develop a model for weighted hypernetworks that couples the establishment of new hyperedges and nodes and the weights' dynamical evolution. We obtain the stationary average hyperdegree and hyperstrength distribution by using the hyperdegree distribution of the hypernetwork model with attractiveness, respectively. In particular, the model yields a nontrivial time evolution of nodes' properties and scale-free behavior for the hyperdegree and hyperstrength distribution. It is expected that our work may give help to the study of the hypernetworks in real-life systems.

## 1. Introduction

When Watts and Strogatz revealed the small-world property of complex networks, and Barabási and Albert discovered scaling in random networks, different kinds of complex networks have attracted great attention from scientists since the late 20th century. A complex network is a graph with non-trivial topological features—features that do not occur in simple graphs such as lattices or random graphs but often occur in graphs modelling real systems. Since then, studies of complex networks are undertaken in many disciplines including mathematics, physics, computer science, biology, social science, economics. Complex network models have been used to study different networks in our life such as protein-protein interaction networks[1], food chain networks[2], transportation networks[3] and large-scale grid networks, economic networks and social networks[4,5]. Through the past decade, scientists have constructed various kinds of models to describe the characteristics of complex networks and proposed many analysis methods to model and optimize networks in real life[6]. Actually, the theoretical studies on complex networks are now making a transition from an original way to a more systematic way.

---

[*] To whom correspondence should be addressed. E-mail: phd5816@163.com

However, some real-life systems are hard to be depicted by complex networks. In many cases the use of complex networks does not provide complete information of the investigated systems. Due to the complication of real-world networks, the common simple graphs are not suitable for networks owning the different kinds of nodes. For example, in author collaboration networks[7], complex networks can only represent the situation that two authors co-work in a paper, while whether there are more than two coauthors linked together cannot be reflected. Many nodes in real-life networks have two or more properties, while nodes in complex networks should maintain homogeneity. For example, nodes in the supply chain[8] obtain different categories including manufacturers, consumers, etc, nodes in the grid network also share different characters including power substations and consumers, simple graphs are not able to represent such systems. Ecological networks are normally represented by competition graph in which we can only know two species competing for their common prey. This kind of graph fails to provide the information about whole groups of species with a particular prey. Competition hypergraph was proposed to yield a more complete description in which nodes denote species and hyperedges denote sets of species having the same prey. In the case of chemical reaction networks nodes and hyperedges are defined as chemical compounds and reactions, respectively. Since chemical reaction is a process containing a set of chemical compounds, substrates, and more than one product, so hypernetwork representation is indispensable[9]. In order to take multi-protein complexes into account a hypergraph is used to represent protein complex networks. In this representation nodes denote protein and hyperedges represent complexes. Only in this way can information about proteins and common protein membership in complexes be taken into account[10]. Although some real-life systems can be represented by bipartite graphs or tripartite graphs, their properties such as small-world, robustness cannot be studied. And the application of measures as node degrees and clustering coefficients to these systems will show differences between these measures for hypernetworks and bipartite graphs.

The emergence of hypernetworks offered a new study method for real-world networks above, and the new concept has been gaining more and more interest in the last years. Bonacich et al. used additional characteristics to extend eigenvector centrality for hypergraphs representing social networks. Park et al.[11] applied the concept of hypergraph theory in cell bio-molecular system, and found that hypergraph structure is very helpful in discovering the building blocks of higher-order interaction of multiple variables, and they also applied the hypergraph model in analysis of mi-croarray data for cancer diagnosis. Akram et al. [12] developed a different application of hypergraphs. They combined intuitionistic fuzzy theory with the hypergraph concept and defined several intuitionistic fuzzy structures which are more flexible than classic models. Zhang et al.[13] built a hypernetwork model of associative memory based on an undirected hypergraph of weighted edges.

Elena et al.[14] used hypergraph theory to study molecular structures of compounds and distinguished these structures by their different topology indices. Wang et al.[15] built a dynamic evolution model for uniform hypernetworks according to growth and preferential attachment mechanisms, in which a new batch of nodes together with one existing node formed one hyperedge in the hypernetwork, and gradually formed the final hypernetwork. Hu et al.[16] proposed another type of dynamic evolution model for uniform hypernetwork. The growth and preferential attachment mechanisms of the model is the same as those of Wang's model, but each time step there will be only one newly added node. Guo and Zhu[17] develop a unified model for uniform hypernetworks and complex networks. Guo and Suo[18] also develop the hypernetwork model with the brand effect and competitiveness. Tian et al.[19] studied the public option intervening and guiding on network based on hypernetwork point of view. Although a few of evolving models in hypernetworks have been proposed based on uniform growth, hypernetworks may have huge potential applications in practical systems.

The above models are unweighted hypernetworks. The purpose of the current work is to extend concept of evolving networks to non-uniform hypernetworks. We propose an attractiveness model of non-uniform hypernetworks and a weighted hypernetwork model to discribe real-life systems better. We obtain the stationary average hyperdegree distribution of the non-uniform hypernetwork by using Poisson process theory and a continuous technique, and theoretically and numerically investigate the hyperdegree distribution. We find that the hyperdegree and hyperstrength distribution of the weighted hypernetwork can be obtained directly from the results of the hypernetwork model with attractiveness.

## 2. Non-uniform hypernetwork with attractiveness

Above hypernetwork models don't take a fact into account that the number of nodes encircled by a new hyperedge is not fixed, however, the number of new nodes entering into the network or previously existing nodes selected at each time step may not be the same. For instance, a blogger often links more than two old blogs in new blogs. In these situations simple uniform hypernetwork model can't provide complete information of the real-life systems. For convenience, the definition of hypernetworks is seen Ref. [18].

A non-uniform hypernetwor model with attractiveness is defined as follows: (i) The network starts from an initial seed of $m_0$ nodes and a hyperedge containing $m_0$ nodes. Suppose that nodes arrive to the system according to a Poisson process $N(t)$ with rate $\lambda$. Each node entering the network is tagged with its own attractiveness $a$. At time $t$, $\eta_{N(t)}$ is the positive integer that is sampled from the population with probability mass function $f(n)$ and $m_1 = \sum_n nf(n)$ is finite, $\xi_{N(t)}$ is the positive integer that is sampled from the population with probability mass function $g(n)$ and $m_2 = \sum_n ng(n)$ is finite. If a new batch of $\eta_{N(t)}$ nodes is added to the network at time $t$, the $\eta_{N(t)}$ new nodes and $\xi_{N(t)}$ previously existing nodes are encircled by a new hyperedge, totally $m$ ($mm_2 \leq m_0$) new hyperedges are constructed with no repetitive hyperedges. (ii) At time t, the probability that a new node will connect to the *j*th node of the *i*th batch, is proportional to the hyperdegree $k_{ij}^h(t)$ and attractiveness $a$, such that

$$\Pi(k_{ij}^h(t)) = \frac{k_{ij}^h(t) + a}{\sum_{ij}(k_{ij}^h(t) + a)}. \tag{1}$$

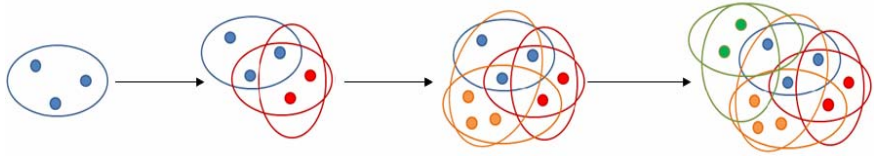

**Fig. 1**: Schematic illustration of the non-uniform hypernetwork evolving process

$t_n$ denotes the time when the *n*th batch of nodes enters into the network. $k_{ij}^h(t)$ denotes the hyperdegree of the *j*th node of the *i*th batch. Supposing that $k_{ij}^h(t)$ is a continuous real-valued variable which is proportional to probability $\Pi(k_{ij}^h)$. Consequently, $k_{ij}^h(t)$ satisfies the dynamical equation by using continuous technique.

$$\frac{\partial k_{ij}^h(t)}{\partial t} = m\xi\lambda \frac{k_{ij}^h(t)+a}{\sum_{ij}(k_{ij}^h(t)+a)} \quad (2)$$

where $\xi$ is a random variable having the distribution $G(n)$.

Since the arrival process of nodes $N(t)$ is a Poisson process, by the Poisson process theory, we know $E[N(t)] = \lambda t$

Since $\sum_{ij}(k_{ij}^h + a) \approx \sum_i m(\eta_i + \xi_i) + a\sum_i \eta_i = \lambda tm(m_1+m_2) + \lambda tam_1$, therefore,

$$\frac{\partial k_{ij}^h}{\partial t} = \frac{m\xi(k_{ij}^h(t)+a)}{(m(m_1+m_2)+am_1)t}. \quad (3)$$

The solution of this equation, with the initial condition that the *i*th batch node at its introduction has $k_{ij}^h(t_i) = m$ is

$$k_{ij}^h(t) = (m+a)\left(\frac{t}{t_i}\right)^{\frac{m\xi}{m(m_1+m_2)+am_1}} - a, \quad (4)$$

From Eq.(4), we get

$$P(k_{ij}^h(t) \geq k) = P\left(t_i \leq \left(\frac{m+a}{k+a}\right)^{\frac{m(m_1+m_2)+am_1}{m\xi}} t\right) \quad (5)$$

Notice that the node arrival process is the Poisson process having rate $\lambda$, therefore the time $t_i$ follows a gamma distribution with parameter $(i, \lambda)$:

$$P(t_i \leq x) = 1 - e^{-\lambda x} \sum_{l=0}^{i-1} \frac{(\lambda x)^l}{l!}$$

Thus,

$$P(t_i \leq (\frac{m+a}{k+a})^{\frac{m(m_1+m_2)+am_1}{m\xi}} t) = 1 - e^{-\lambda t \, (\frac{m+a}{k+a})^{\frac{m(m_1+m_2)+am_1}{m\xi}}} \sum_{l=0}^{i-1} \frac{1}{l!} (\lambda t \, (\frac{m+a}{k+a})^{\frac{m(m_1+m_2)+am_1}{m\xi}})^l . \quad (6)$$

Substituting Eq. (6) into Eq. (5) yields

$$P(k_{ij}^h(t,\xi) \geq k) = 1 - e^{-\lambda t \, (\frac{m+a}{k+a})^{\frac{m(m_1+m_2)+am_1}{m\xi}}} \sum_{l=0}^{i-1} \frac{1}{l!} (\lambda t \, (\frac{m+a}{k+a})^{\frac{m(m_1+m_2)+am_1}{m\xi}})^l . \quad (7)$$

Next we verify that stationary average hyperdegree distributions exist in the hypernetwork. From Eq. (7), we obtain

$$P(k_{ij}^h(t,\xi) = k) \approx \frac{\partial P(k_{ij}^h(t,\xi) < k)}{\partial k}$$

$$= \lambda t \frac{m(m_1+m_2)+am_1}{m(m+a)\xi} (\frac{m+a}{k+a})^{\frac{m(m_1+m_2)+am_1}{m\xi}+1} \frac{(\lambda t \, (\frac{m+a}{k+a})^{\frac{m(m_1+m_2)+am_1}{m\xi}})^{i-1}}{(i-1)!} e^{-\lambda t \, (\frac{m+a}{k+a})^{\frac{m(m_1+m_2)+am_1}{m\xi}}} . \quad (8)$$

From Eq. (8), we get the following equation for the stationary average hyperdegree distribution of the hypernetwork,

$$P(k) \approx \frac{m(m_1+m_2)+am_1}{m(m+a)} \sum_{\xi} \frac{1}{\xi} (\frac{m+a}{k+a})^{\frac{m(m_1+m_2)+am_1}{m\xi}+1} g(\xi) \quad (9)$$

When $\eta = m_1, \xi = m_2$, from Eq.(9), we have

$$P(k) = \frac{m(m_1+m_2)+am_1}{m_2 m(m+a)} (\frac{m+a}{k+a})^{\frac{m(m_1+m_2)+am_1}{mm_2}+1} \quad (10)$$

Eq. (10) exhibit the scale-free property of the hypernetwork, and the hyperdegree distribution behaves as $P(k) \propto k^{-\gamma}$ where

$$\gamma = \frac{m(m_1+m_2)+am_1}{mm_2} + 1 \quad (11)$$

In the following simulation, take $m_0=10$, $m=2$ and $a=1$. The simulation results are showed from Figure 2 to Figure 3 in double-logarithmic axis. As the figures show, the theoretical prediction of the hyperdegree distribution is in good agreement with the simulation results.

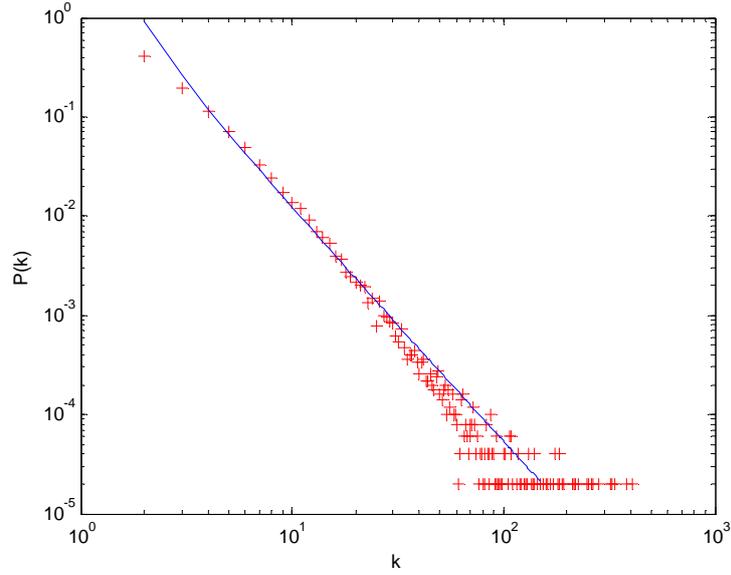

**Fig. 2**: The simulation of the non-uniform hypernetwor model. $N=100000$, $\eta_{N(t)}$ is random selected from 1~3, $\xi_{N(t)}$ is random selected from 1~5. + denotes the simulation result, the line denotes theoretical prediction.

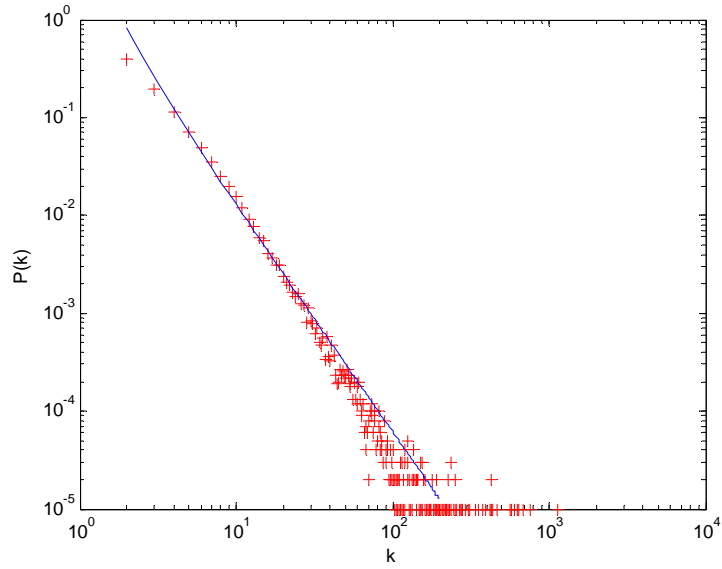

**Fig. 3**: The simulation of the non-uniform hypernetwor model. $N=150000$, $\eta_{N(t)}$ is random selected from 1~2, $\xi_{N(t)}$ is random selected from 1~4. + denotes the simulation result, the line denotes theoretical prediction.

## 3. Weighted hypernetworks

In the BBV (Barrat-Barthelemy-Vespignani) model proposed by Barrat, et al. nodes enter into the network one by one and the edges formed by one new added node and one old node. This model can only represent relations between a pair of nodes[20,21]. However, edges in many real-world systems should involve information such as cooperation, trade or interaction among more than two actors. For instance, the authors collaborating network[2] is a weighted hypernetwork. The weight of hyperedges should be the number of papers cooperated by co-authors. In the airline networks the weight of edges was used to represent passenger flow volume. In the trade networks the weight of edges was used to represent total trade between countries [22]. In transportation networks, metro lines are always added more than one node at each time step. These networks are different from simple weighted networks. The paper proposes a model of weighted evolving hypernetworks to describe the weighted hyperedge growth caused by batches of newly added nodes. We also obtain the theoretical analyses result.

The mathematical definition of the weighted hypergraph is as follows. Let $V = \{v_1, v_2, \cdots, v_n\}$ be a finite set, and let $E_i = \{v_{i_1}, v_{i_2}, \cdots, v_{i_k}\}$ $(v_{i_j} \in V, j = 1,2,\cdots,k)$, $E^h = \{E_1, E_2, \cdots, E_L\}$ be a family of subsets of $V$, $w$ is a map from $E^h$ into real number set $R$, denoted by $w_k = w(E_k), W = \{w_1, w_2, \cdots w_L\}$ The triple $(V, E^h, W)$ is called a weighted hypergraph. The elements in $V$ are called a node set, and $E_i$ $(1,2,\cdots,L)$ is a set of non-empty subsets of $V$ called a hyperedge set. In a weighted hypergraph, two nodes are said to be adjacent if there is a hyperedge that contains both of these nodes. Two hyperedges are said to be adjacent if their intersection is not empty. If $|V|$ and $|E^h|$ are finite, $H$ is a finite weighted hypergraph. If $|E_i| = u$ $(i = 1,2,\cdots,L)$, $H = (V, E^h)$ is an $u$-uniform weighted hypergraph. If $|E_i| = 2$, $(i = 1,2,\cdots,L)$, $H = (V, E^h)$ degrades to a weighted network.

Based on the above definitions, we can give mathematical definition of the weighted hypernetwork. Suppose $\Omega = (V, E^h, W)$ is a finite weighted hypergraph and $G$ is a map from $T = [0,+\infty)$ into $\Omega$; for any given $t \geq 0$, $G(t) = (V(t), E^h(t), W(t))$ is a finite weighted hypergraph. The index *t* is often interpreted as time. A weighted hypernetwork $\{G(t), t \in T\}$ is a collection of weighted hypergraphs. The hyperdegree of $v_i$ is defined as the number of hyperedges that connect to node $v_i$. For the hyperedges that connected to $v_i$, the sum of their hyperedge weight is called the hyperstrength of $v_i$. The definition of the weighted hypernetwork model is based on two coupled mechanisms: the topological growth and the weights' dynamics. The weighted hypernetwork model is defined as follows :

(ⅰ) *Growth*: The network starts from an initial seed of $m_0$ nodes and a hyperedge containing $m_0$ nodes, and the hyperedge is assigned weight $w_0$. Suppose that nodes arrive to the system according to a Poisson process with rate $\lambda$. If $m_1$ new nodes arrive to the network at time $t$, one new hyperedge is formed by these new nodes and $m_2$ ($mm_2 \leq m_0$) previously existing nodes, totally $m$ new hyperedges are constructed with no repetitive hyperedges.

(ⅱ) *Hyperstrength driven attachment*: The new batch nodes preferentially choose nodes with larger hyperstrength, i.e., the probability that the new batch nodes will connect to previously existing node $v_{ij}$ of the *j*th node of the *i*th batch is proportional to the hyperstrength $s_{ij}^h(t)$ of node $v_{ij}$, such that

$$\prod(k_{ij}^h(t)) = \frac{s_{ij}^h}{\sum_{i,j} s_{ij}^h}, \tag{12}$$

where $k_{ij}^h(t)$ is the hyperdegree of node $v_{ij}$, $s_{ij}^h = \sum_{k|v_{ij} \in E_k} w_k$ is the hyperstrength of node $v_{ij}$.

(iii) *Weights' dynamics*: The weight of each new hyperedge is initially set to a given value $w_0$. A new hyperedge of node $v_{ij}$ will trigger only local rearrangements of weights on the previously existing neighbors $v_{rl} \in N_{v_{ij}}$, where $N_{v_{ij}}$ represents the neighbors of $v_{ij}$, according to the simple rule

$$w_k \to w_k + \Delta w_k , \tag{13}$$

where $\Delta w_k = \delta \frac{w_k}{s_{ij}^h}$, $v_{ij}, v_{rl} \in E_k$, $\delta$ is defined as updating coefficient and $\delta = const$.

When a new batch arrive to the system, an already present node $v_{ij}$ can be affected in two ways: (a) It is chosen with probability (12) to be connected to the batch of new nodes, then its hyperdegree increases by 1, and its hyperstrength by $w_0 + \delta$. (b) One of its neighbors $v_{rl} \in N_{v_{ij}}$ is chosen to be connected to the batch of new nodes, then the hyperdegree of $v_{ij}$ is not modified, but $w_k$ is increased according to the rule Eq. (13), and thus $s_{ij}$ is increased by $\delta \frac{w_k}{s_{rl}^h}$. This dynamical process modulated by the respective occurrence probabilities $\frac{s_{ij}^h(t)}{\sum_{i,j} s_{ij}^h(t)}$ and $\frac{s_j^h(t)}{\sum_{r,l} s_{rl}^h(t)}$ is thus described by the following evolution equations for $s_{ij}^h(t)$ and $k_{ij}^h(t)$:

$$\frac{ds_{ij}^h}{dt} = mm_2(w_0 + \delta)\frac{s_{ij}^h}{\sum_{i,j} s_{ij}^h} + \sum_{k|v_{ij} \in E_k, v_{rl} \in N_{v_{ij}}} mm_2 \frac{s_{rl}^h}{\sum_{r,l} s_{rl}^h} \delta \frac{w_k}{s_{rl}^h} \qquad (14)$$

$$\frac{dk_{ij}^h}{dt} = mm_2 \frac{s_{ij}^h}{\sum_{i,j} s_{ij}^h}, \qquad (15)$$

Since $\sum_{k|v_{ij} \in E_k, v_{rl} \in N_{v_{ij}}} mm_2 \frac{s_{rl}^h}{\sum_{r,l} s_{rl}^h} \delta \frac{w_k}{s_{rl}^h} = mm_2 \delta \frac{s_{ij}^h}{\sum_{r,l} s_{rl}^h}$, therefore, the following is obtained

$$\frac{ds_{ij}^h}{dt} = mm_2(w_0 + 2\delta)\frac{s_{ij}^h}{\sum_{i,j} s_{ij}^h} \qquad (16)$$

Substituting Eq. (15) into Eq. (16) yields

$$\frac{ds_{ij}^h}{dt} = (w_0 + 2\delta)\frac{dk_{ij}^h}{dt}$$

Since node $v_{ij}$ arrives to the system at time $t_i$, we have $k_{ij}^h(t_i) = m$ and $s_{ij}^h(t_i) = mw_0$, then the above equation is integrated from $t_i$ to $t$, the following is obtained

$$s_{ij}^h = (w_0 + 2\delta)k_{ij}^h - 2\delta m, \qquad (17)$$

and probability (12) is modified as follows:

$$\Pi(k_{ij}^h(t)) = \frac{k_{ij}^h - \frac{2\delta}{(w_0 + 2\delta)}m}{\sum_{i,j}\left[k_{ij}^h - \frac{2\delta}{(w_0 + 2\delta)}m\right]} \quad (18)$$

By comparing probability (18) and probability (1), it can be inferred that the attractiveness of the hypernetwork model is as follows

$$a = -\frac{2\delta}{w_0 + 2\delta}m, \quad (19)$$

The probability of the preferential attachment in this model can be modified as $\Pi(k_{ij}(t)) = \frac{k_i^h + a}{\sum_j (k_j^h + a)}$, which is in accord with that of the evolving hypernetwork model with attractiveness. Substituting Eq. (19) into Eq. (10) yields the stationary average hyperedegree distribution of the weighted hypernetwork

$$P(k) \approx \frac{w_0 + 2\delta}{mw_0}(\frac{m_1 w_0}{m_2(w_0 + 2\delta)} + 1)(\frac{mw_0}{(w_0 + 2\delta)k - 2\delta m})^{\frac{m_1 w_0}{m_2(w_0 + 2\delta)} + 2} \quad (20)$$

Moreover, from Eq.(20), the hyperdegree distribution of the weighted hypernetwork behaves as $P(k) \propto k^{-\gamma}$ where

$$\gamma = 2 + \frac{m_1}{m_2}\frac{w_0}{w_0 + 2\delta} \quad (21)$$

Therefore, the hyperdegree distribution of the weighted hypernetwork can be obtained directly from the results of the evolving hypernetwork model with attractiveness.

When $\eta = m_1, \xi = m_2$, from Eq.(7), Eq.(17) and Eq.(19), we have

$$P(s_{ij}^h(t) < x) = e^{-\lambda t\left(\frac{w_0 m}{x}\right)^{1+\frac{m_1}{m_2}(\frac{w_0}{w_0+2\delta})}} \sum_{l=0}^{i-1}\frac{1}{l!}(\lambda t\ (\frac{w_0 m}{x})^{1+\frac{m_1}{m_2}(\frac{w_0}{w_0+2\delta})})^l. \quad (22)$$

Hence, the density function of $s_{ij}$ is as follows

$$f_{s_{ij}}(x) \approx \lambda t(1 + \frac{m_1}{m_2}(\frac{w_0}{w_0+2\delta}))\frac{(w_0 m)^{1+\frac{m_1}{m_2}(\frac{w_0}{w_0+2\delta})}}{x^{2+\frac{m_1}{m_2}(\frac{w_0}{w_0+2\delta})}}e^{-\lambda t\left(\frac{w_0 m}{x}\right)^{1+\frac{m_1}{m_2}(\frac{w_0}{w_0+2\delta})}}\frac{(\lambda t\ (\frac{w_0 m}{x})^{1+\frac{m_1}{m_2}(\frac{w_0}{w_0+2\delta})})^{i-1}}{(i-1)!}. \quad (23)$$

Then, the density function $f(x)$ of the stationary average hyperstrength distribution can be deduced from Eq. (23) as follows:

$$f(x) \approx (1+\frac{m_1}{m_2}(\frac{w_0}{w_0+2\delta}))(w_0 m)^{1+\frac{m_1}{m_2}(\frac{w_0}{w_0+2\delta})}(\frac{1}{x})^{2+\frac{m_1}{m_2}(\frac{w_0}{w_0+2\delta})}. \quad (24)$$

Evidently, from Eq.(24), we know that the stationary average hyperstrength distribution of the weighted hypernetwork is a power-law distribution.

## 4. Conclusion

The paper proposes the non-uniform hypernetwork model with attractiveness and the evolving model of weighted hypernetworks. In the non-uniform hypernetwork the new batch of size and the number of randomly selected existing nodes are random variables, respectively. We obtain a formula of the stationary average hyperdegree distribution of the non-uniform hypernetwork. The analysis is verified with numerical simulation results. When the model degenerates to an uniform hypernetwork, a power-law behavior with exponent $\gamma = 2 + \frac{m_1}{m_2}(1+\frac{a}{m})$ is displayed. The weighted hypernetwork takes the topological growth and the weights' dynamics mechanisms into account. We find that the weighted hypernetwork is a special case of the non-uniform hypernetwork model with attractiveness. The study of hypernetwork is necessary for the future multidisciplinary research. The application of hypernetworks in real life system is worth the further investigation. We expect that the result in this paper can accelerate investigations of hyperneworks. In this perspective, the present model appears as a general starting point for the realistic modeling of weighted hypernetworks.

### ACKNOWLEDGMENTS

The authors acknowledge support from the National Natural Science Foundation of China (Grant No. 71571119), the Shanghai First-class Academic Discipline Project, China (Grant No. S1201YLXK), and supported by the Hujiang Foundation of China (Grant No. A14006).